\documentclass[a4paper]{aa}
\usepackage{txfonts}
\usepackage{colortbl}
\usepackage{etoolbox}
\usepackage{graphicx}
\usepackage{amsmath}
\usepackage{hyperref}
\newcommand{\etal}{{et al}\/.}
\newcommand\rd{{\rm d}}

\hypersetup{colorlinks=true,linkcolor=blue,citecolor=blue,filecolor=blue,urlcolor=blue}
\begin{document}
\title{The contribution of discrete sources to the sky temperature at 144 MHz}
\titlerunning{The sky temperature at 144 MHz}
\author{
  M.J.\ Hardcastle\inst{1}\thanks{m.j.hardcastle@herts.ac.uk} \and
  T.W.\ Shimwell\inst{2} \and
  C.\ Tasse\inst{3,4} \and
  P.N. Best\inst{5} \and
  A.\ Drabent\inst{6} \and
  M.J.\ Jarvis\inst{7,8} \and
  I. Prandoni\inst{9} \and
  H.J.A.~R\"ottgering\inst{10} \and
  J.\ Sabater\inst{5} \and
  D.J.\  Schwarz\inst{11}}
\institute{
Centre for Astrophysics Research, University of
       Hertfordshire, College Lane, Hatfield AL10 9AB, UK \and
ASTRON, Netherlands Institute for Radio Astronomy, Oude
  Hoogeveensedijk 4, Dwingeloo, 7991 PD, The Netherlands \and
GEPI, Observatoire de Paris, CNRS, Universit\'e Paris Diderot, 5
  place Jules Janssen, 92190 Meudon, France \and
Department of Physics \& Electronics, Rhodes University, PO Box
  94, Grahamstown, 6140, South Africa \and
SUPA, Institute for Astronomy, Royal Observatory, Blackford Hill,
  Edinburgh, EH9 3HJ, UK \and
Th\"uringer Landessternwarte, Sternwarte 5, 07778 Tautenburg, Germany \and
Astrophysics, Department of Physics, Keble Road, Oxford, OX1 3RH, UK \and
Department of Physics \& Astronomy, University of the Western
  Cape, Private Bag X17, Bellville, Cape Town, 7535, South Africa \and
INAF-Istituto di Radioastronomia, Via P. Gobetti 101, 40129
Bologna, Italy \and
Leiden Observatory, Leiden University, PO Box 9513, 2300 RA Leiden, The Netherlands \and
Fakult\"at f\"ur Physik, Universit\"at Bielefeld, Postfach 100131, 33501 Bielefeld, Germany}

\abstract{In recent years, the level of the extragalactic radio
  background has become a point of considerable interest, with some
  lines of argument pointing to an entirely new cosmological
  synchrotron background. The contribution of the known discrete
  source population to the sky temperature is key to this discussion.
  Because of the steep spectral index of the excess over the cosmic
  microwave background, it is best studied at low frequencies where
  the signal is strongest. The Low-Frequency Array (LOFAR) wide and
  deep sky surveys give us the best constraints yet on the
  contribution of discrete extragalactic sources at 144 MHz, and in
  particular allow us to include contributions from diffuse,
  low-surface-brightness emission that could not be fully accounted
  for in previous work. We show that, even with these new data, known
  sources can still only account for around a quarter of the estimated
  extragalactic sky temperature at LOFAR frequencies.}

\maketitle

\section{Introduction}
\label{sec:intro}
There is a good deal of interest in the level and origin of the
extragalactic radio background, primarily as a result of
high-frequency work using the ARCADE-2\footnote{ARCADE stands for
  `Absolute Radiometer for Cosmology, Astrophysics, and Diffuse
  Emission'. Although the
  evidence for the background discussed in this section pre-dates the ARCADE-2 result, it was
  the ARCADE-2 results that gave it its current prominence and so we
  refer to it as the ARCADE-2 background throughout the paper.} bolometer \citep{Fixsen+11}.
This work detected an excess over the cosmic microwave background
radiation (CMB) at high radio frequencies
after subtraction of a model of Milky Way emission, which was shown to
connect well with previously reported estimates of the extragalactic
sky temperature at gigahertz frequencies and below.
Although an extragalactic background is expected from summing the
effects of the many discrete extragalactic radio sources that are known to be
present more or less isotropically across the sky \citep{Bridle67},
the level of the ARCADE-2 background was much higher than could easily
be explained by such an analysis \citep{Vernstrom+11,Condon+12}. An
overview of the state of the subject is given by \cite{Singal+18}; as
they point out, if the background seen by ARCADE-2 and numerous
low-frequency experiments is genuinely extragalactic in origin, its
nature is one of the most interesting unsolved problems in
astrophysics, since the sources responsible for the background would
have to be very faint, non-thermal (given the spectral index of the
background), and much more numerous than known galaxies. The existence
of this background has motivated more exotic physical and
astrophysical explanations, such as a role for radio-loud primordial
black holes \citep*[e.g.][]{Ewall-Wice+20}.

It is challenging to make direct measurements of the (presumably
isotropic) extragalactic background emission at low frequencies
because of the strong effects of synchrotron emission from the Milky
Way. Particularly interesting is the frequency range around 150 MHz,
which is well studied both because of its traditional use in early
radio astronomy and because radio detections of the epoch of
reionisation (EoR) are expected in this frequency range. Experiments
searching for the signature of Cosmic Dawn, such as EDGES\footnote{Experiment to Detect the Global EoR Signature} \citep{Bowman+18},
also need to take account of the low-frequency extragalactic discrete background.

Early work in mapping the low-frequency sky temperature by
\cite{Turtle+Baldwin62} gave a minimum sky temperature at 178 MHz of
80 K, which sets an upper limit on the isotropic component, modulo
large uncertainties. \cite{Bridle67} argued that the temperature of
the extragalactic isotropic component does not exceed $30 \pm 7$ K at
178 MHz, based on a combination of several low-frequency surveys.
However, the ARCADE-2 result, which at its low-frequency end is based
on fits to data at 22, 45, 408, and 1400 MHz and does not use the early
low-frequency surveys, predicts a much higher background of 110 K at
178 MHz. Further subsequent low-frequency work seems to confirm the
original ARCADE-2 result \citep{Dowell+Taylor18}, but again at
frequencies of 40-80 MHz. No direct measurement of this background at
100-200 MHz has been published in recent years. As \cite{Singal+18}
point out, a wide-area survey with good zero level calibration in this
frequency range would potentially confirm or rule out the current
inferred background levels and, in that context, it is interesting to
ask what constraints we have on the known extragalactic contributions
at those frequencies.

In this paper we use the wide and deep Low-Frequency Array (LOFAR) surveys currently being
carried out at frequencies between 120 and 168 MHz to revisit the
question of what fraction of the radio background can be produced by
known, catalogued discrete sources. We describe the LOFAR data in
Section 2 and our analysis of them in Section 3. Our discussion and
conclusions are in Section 4.

\section{Data}

To assess the contribution of LOFAR sources to the extragalactic
background we use both wide and deep-field data from the LOFAR
northern-sky survey LoTSS\footnote{LoTSS is the LOFAR Two-Metre Sky
  Survey: see \url{https://lofar-surveys.org/}.} \citep{Shimwell+17}. LoTSS images using the
Dutch baselines have relatively high resolution (6 arcsec) at low
frequency (central frequency of 144 MHz, typically with 48 MHz of
bandwidth) but crucially, unlike comparable Very Large Array (VLA) surveys, are also
sensitive to extended emission on a wide range of scales, which may
make a significant contribution to the background levels
\citep{Vernstrom+15}. Because of the low selection frequency, the
LOFAR source population is completely dominated by non-thermal
emitters, with a typical spectral index expected to be comparable to
that estimated for the ARCADE-2 excess
(\citealt{Mauch+13,Hardcastle+16}; Sabater \etal\ in preparation).

The wide-area data are images and radio
source catalogues from the planned second data release of LoTSS (DR2),
which will cover 5,700 degrees of the Northern sky. The DR2
data were processed with version 2.2 of the standard Surveys Key
Science Project
pipeline\footnote{\url{https://github.com/mhardcastle/ddf-pipeline}},
as described by \cite{Shimwell+19} and Tasse \etal\ (2020, in prep.). This pipeline carries out
direction-dependent calibration using {\sc killMS}
\citep{Tasse14,Smirnov+Tasse15} and imaging is done using {\sc
  DDFacet} \citep{Tasse+18}. DR2 uses a new self-calibration and
imaging strategy which improves the sensitivity to extended emission
relative to the first data release, DR1; the new strategy was
described briefly by \cite{Shimwell+19} and is discussed in more
detail by Tasse et al.\ (2020). The individual pointings are on an
overlapping tiling of the sky and so we combine the observations of
each pointing with those of its immediate neighbours with appropriate
weighting in the image plane, again as described by
\cite{Shimwell+19}, to make a set of `mosaics'. Source catalogues are
then extracted for each mosaic and a non-redundant
catalogue is generated by choosing the catalogue entry derived from
the mosaic with the best central astrometry
(as estimated during the pipeline processing) in each region of overlapping sky
area.

We cropped the version 1.00 DR2 catalogue released to the LoTSS
collaboration in March 2020 to make a
catalogue of 1,913,117 sources covering 1970 square degrees in the
area $130^\circ < \alpha < 250^\circ$, $40^\circ < \delta < 65^\circ$.
This region is centred on the HETDEX field\footnote{HETDEX is the
  Hobby-Eberly Dark Energy Experiment. LoTSS's early observations
  covered the HETDEX Spring field.} that was the target of DR1,
but covers nearly 5 times the sky area and of course is more sensitive
and better calibrated than the publicly available DR1 data. The
observations of the individual pointings that make up this region
(hereafter the `DR2 13-h field' to distinguish it from the much larger
area to be covered by DR2 itself) were reduced to obtain roughly
uniform sensitivity over 316 mosaics. All the sky area used here is at
relatively high Galactic latitude, $35^\circ < \ell < 75^\circ$, but
in any case the shortest baseline cut of 100 m imposed in imaging
means that we are insensitive to Galactic diffuse emission (or any
emission on angular scales $\ga 1^\circ$: we return to this point below). The
typical rms noise across the whole field is $\sim 70$ $\mu$Jy beam$^{-1}$ at
6-arcsec resolution.

LOFAR absolute flux calibration is well known to be problematic. In
DR1 correction of the initial flux calibration was done as part of the
pipeline using a `bootstrap' process described by
\cite{Hardcastle+16}. For DR2 we investigated the quality of this
bootstrap process by crossmatching bright ($>200$ mJy) compact sources
with the 6C catalogue \citep{Hales+88,Hales+90}, which covers all of
the area of our 13-h field. 6C by design is on the flux scale of
\citet*{Roger+73}, since it includes scans over the primary reference
source Cygnus A (hence \cite{Scaife+Heald12} adopt 6C flux
densities without correction). We found that the uncorrected DR2
absolute flux scale is very consistent with the flux scale of 6C, when
the slightly different observing frequencies are taken into account;
however, there was significant residual scatter which could be
improved by scaling the images for each individual pointing using the
ratio of LoTSS to NRAO VLA Sky Survey (NVSS) flux densities. This scaling was applied to the
images used to make our mosaics and we confirm that the overall flux
scale of the 13-h field is consistent with that of the 6C catalogue to
within 2 per cent, with a per field standard deviation around 8 per
cent. Given that the quoted 6C flux scale uncertainties are around 5
per cent this suggests a similar uncertainty on the DR2 flux scale. We
adopt 5 per cent as the fundamental limiting flux scale uncertainty
for the wide-area data in what follows.

In addition to the wide-area data, we used the much smaller LoTSS Deep
Field data release 1 observations of the Bo\"otes,
ELAIS-N1\footnote{ELAIS is the European Large Area {\it ISO} Survey.} and
Lockman Hole areas. These are single LOFAR pointings reduced in
essentially the same manner (see Tasse et al. in prep.\ for the
process used for Bo\"otes and Lockman Hole and Sabater et al in
prep.\ for ELAIS-N1) but with much longer integration times, leading
to nominal central rms noise levels of 30, 23 and 20 $\mu$Jy
beam$^{-1}$ respectively. All of these fields are also at high
Galactic latitude (ELAIS-N1 and Lockman are within the 13-h field,
Bo\"otes is just to the south of it) and so are very comparable to the
DR2 data except in depth. We cropped the three deep-field images at a
primary beam sensitivity factor of 0.87, giving a sky area of roughly
3 deg$^2$ for each field; by giving roughly uniform sensitivity across
the field this simplifies completeness analysis (see below) and a
larger area is not necessary as we have the 13-h field to draw on. 
These three fields have had flux scale corrections applied using
detailed comparison with other radio observations, but are all
broadly consistent with the scale in the DR2 catalogue, which had
previously been corrected to the \cite{Roger+73} flux scale as
described above, with maximum deviation from the DR2 scale of order 10
per cent. For simplicity we adopt the flux scales used in each
catalogue, but given the uncertainties associated with calibrating a
small field we apply a flux scale uncertainty of 10 per cent to these
fields. We note (see
Kondapally et al in prep.) that the optical identification fraction of
the sources in these fields with galaxies or quasars is nearly 100\% ---
thus we can be certain that they represent a predominantly
extragalactic population.

Details of the four fields used and their properties are given
in Table \ref{tab:flist}.

\begin{table}
  \caption{Fields used in the analysis}
  \label{tab:flist}
  \begin{tabular}{lrrr}
    \hline
    Name&Number of sources&Area&RMS noise$^a$\\
    &in area used&(deg$^2$)&($\mu$Jy beam$^{-1}$)\\
    \hline
    DR2 13h&1,913,117&1970.0&62\\
    Bo\"otes&9,167&3.216&39\\
    ELAIS-N1&15,166&2.886&21\\
    Lockman&12,268&3.042&26\\
    \hline
  \end{tabular}
  \vskip 5pt $^a$ RMS noise tabulated is the median island rms
  reported by PyBDSF for the catalogue actually used for the analysis,
  and so differs from the noise levels quoted in the text because it
  depends on the regions of sky used and because it is biased towards
  low-noise regions.
\end{table}

For both the DR2 13-h field and the three cropped deep fields that we
use our main input datasets are radio sky catalogues derived using the
Python Blob Detector and Source Finder ({\sc PyBDSF}) software
\citep{Mohan+Rafferty15}. For the DR2 data these were simply the DR2
v1.0 internal release catalogue with RA and DEC limits applied as
described above, while we extracted our own catalogues from the cropped
deep fields images using the same {\sc PyBDSF} settings as used for
the wide-field mosaics. In particular, {\sc PyBDSF}'s wavelet mode was
used to allow the cataloguing of faint, low-surface-brightness
emission. Throughout the paper we refer to distinct objects classified
by {\sc PyBDSF} as `sources'. We do not make use of any associations
between distinct {\sc PyBDSF} sources, although these are available
for the deep fields (see Kondapally \etal\ in prep.) since they have
not yet been produced for DR2. We do not expect this to have any
effect on our analysis.

\section{Analysis}

In this section we compute the contribution of discrete sources seen
by LOFAR (spanning the range of source flux densities seen in the wide
and deep surveys, and so including sources with flux densities between 100 Jy and 100 $\mu$Jy) to the
total sky temperature at 144 MHz. We begin by constructing the 144-MHz
source counts; we then discuss the origin of incompleteness in the
uncorrected source counts and derive a method for correcting for it,
and finally estimate the total contribution of corrected and uncorrected
counts to the sky temperature.

\subsection{Uncorrected source counts}

\begin{figure*}
  \includegraphics[width=0.48\linewidth]{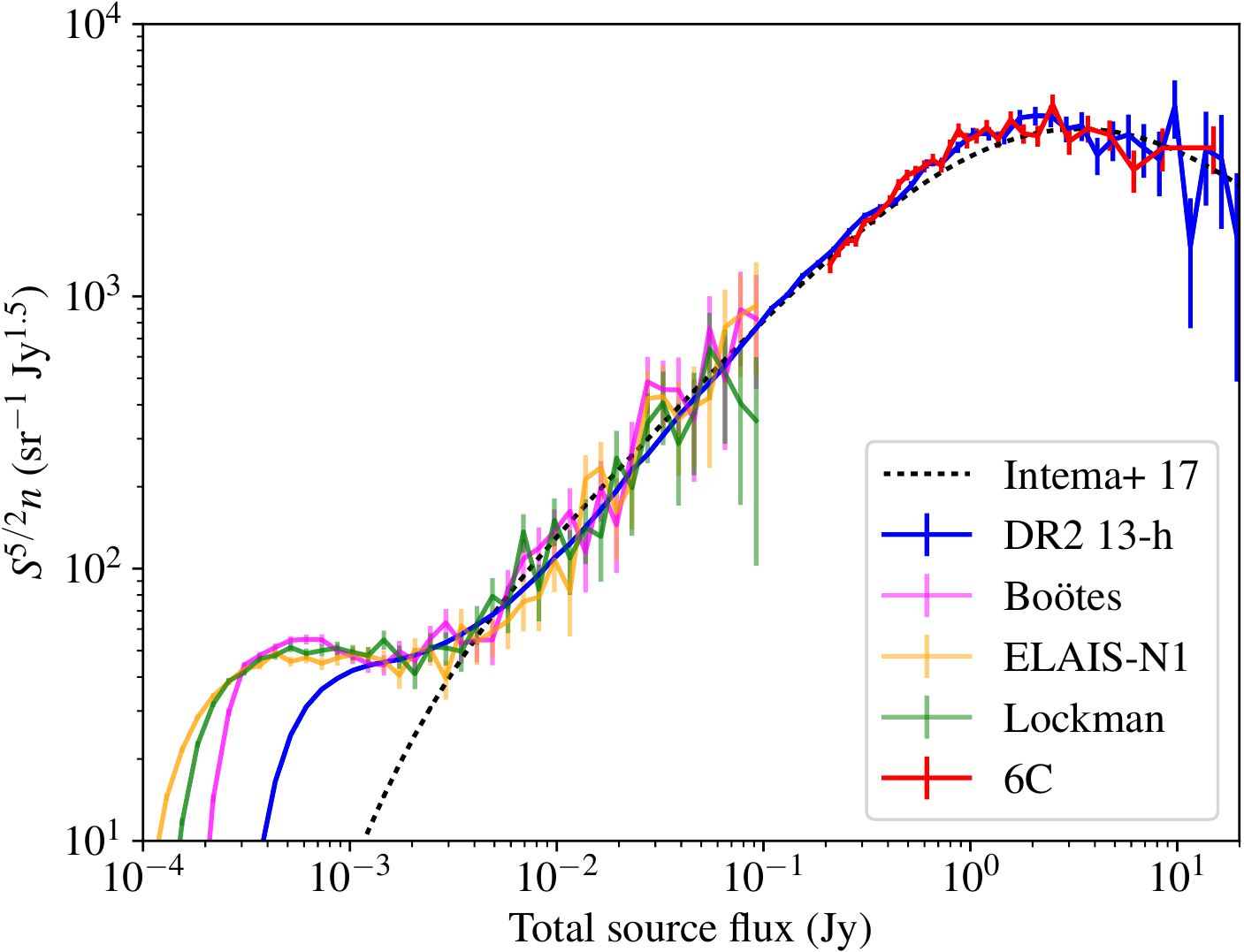}
  \includegraphics[width=0.48\linewidth]{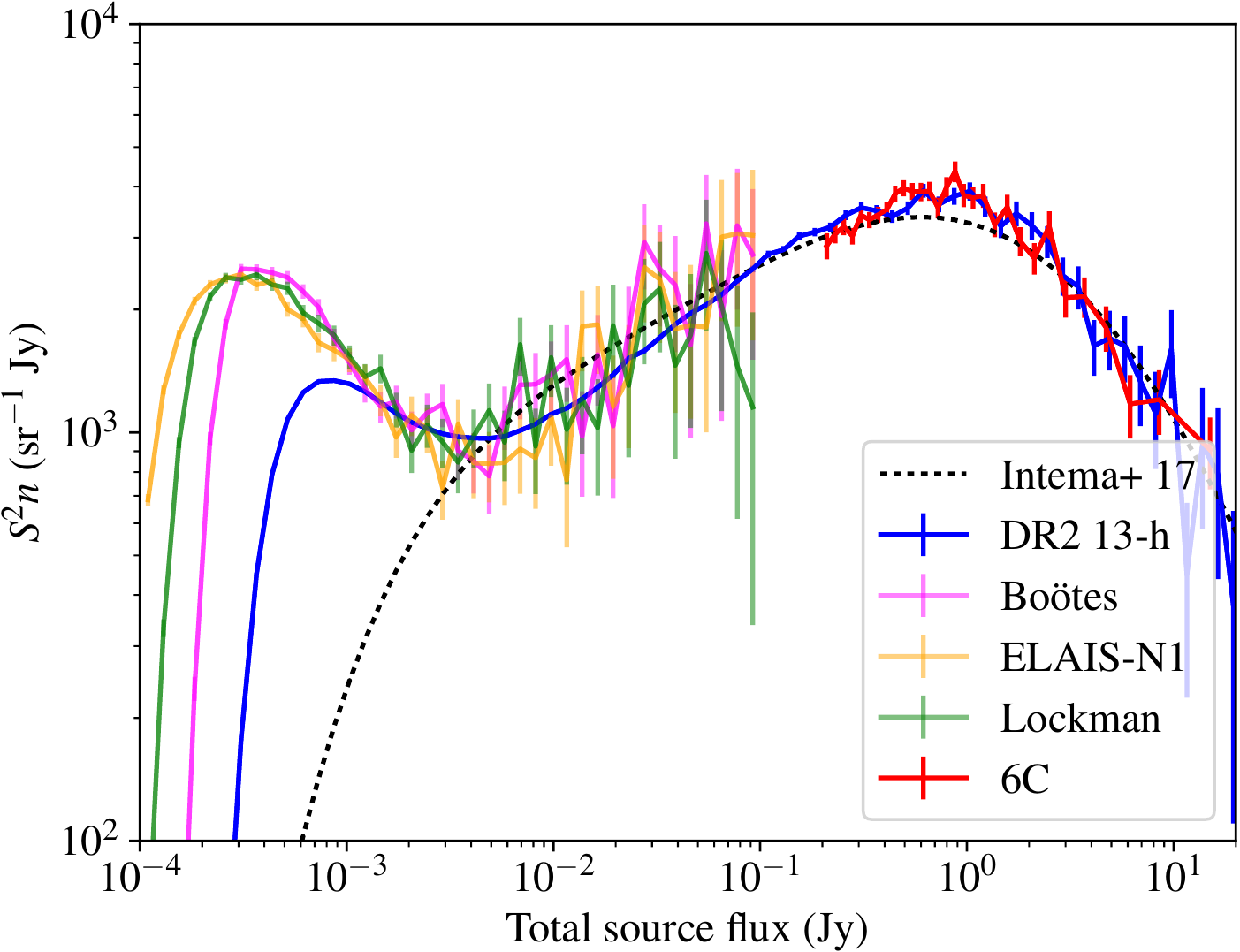}
\caption{Uncorrected source counts for the DR2 13-h field and the three
  LoTSS deep fields. Poissonian error bars are plotted. The left panel shows the conventional Euclidean
  scaling of the differential source counts, $S^{5/2}$; the right
  panel shows a scaling by $S^2$ which emphasises the contribution
  made by each bin to the integrated total flux density per unit area.
  Overplotted are the 6C source counts of \protect\cite{Hales+88} and the
  functional form fitted by \protect\cite{Intema+17} to the combined
  TGSS and literature source counts.}
\label{fig:sourcecounts}
\end{figure*}

Fig.\ \ref{fig:sourcecounts} shows the source counts at 144 MHz for
DR2 and the three deep fields. We see that there is good agreement at
the bright end with the shape and normalisation of the 6C counts of
\cite{Hales+88}, and reasonable agreement with the fitted functional
form of \cite{Intema+17} using counts from the TIFR GMRT Sky Survey
Alternative Data Release (TGSS), which, however, seems systematically low by
around 20 per cent with respect to both LoTSS and 6C at flux densities
around 1 Jy, and of course deviates from observations below the flux
density limit of 5 mJy used by \citeauthor{Intema+17} A crossmatch
between TGSS and 6C suggests a flux scale offset of around 6 per cent
for bright sources (in the sense that TGSS fluxes are systematically
higher than 6C) which may account for some of the discrepancy. The
source counts we present here are generally consistent with both
bright-end and faint-end source counts from LOFAR already presented in
the literature or shortly to be published
\citep{Williams+16,Hardcastle+16,Retana-Montenegro+18,Hale+19,Siewert+20} and, as
the source counts are not the main topic of this paper, we do not
discuss them in more detail here: the reader is referred to the paper
by Mandal
\etal\ (in prep.), who make use of the full area of the deep fields, for discussion of the implications of the deep-field counts.

At flux densities below a few mJy the LOFAR counts start to be
affected by incompleteness, and, as expected, this effect is more
significant for the 13-h field than the three deep fields. From the
point at which the wide-area source counts start to fall below those
for the deep fields we can conservatively estimate that the 13-h field
is affected by incompleteness to some level below about 4 mJy. The
three deep fields, which give much better sampling of the sub-mJy
population, then exhibit cutoffs that are qualitatively consistent
with their different depths. It is noteworthy that the normalisations
of the three deep-field curves differ by a few per cent in the
flux density range 0.5--1.0 mJy. This could be due to either cosmic variance
or residual flux scale uncertainties or a combination of the two. For
the small areas of the deep fields that we use here, with radius $\sim
1^\circ$, cosmic variance can be significant \citep{Siewert+20}, in which case the Poisson errors plotted on the data
points underestimate their true uncertainty. The modelling of
\cite{Heywood+13} suggests that we would expect cosmic variance on its
own to give a few per cent uncertainty on the number
counts at these flux density levels in 3 deg$^2$ areas, so it is
plausible that the differences seen here are dominated by cosmic variance.
If so, the flux scale uncertainty of 10 per cent that we assign to these
fields may be unnecessarily conservative, but we retain it in what follows.

In the right-hand panel of Fig. \ref{fig:sourcecounts} we can see that
the sub-mJy source population reaches levels on an $S^2 n(S)$ plot
comparable to the dominant $\sim 1$ Jy population, and will thus make
a non-negligible contribution to the total integrated flux density.
The power-law source-counts function for these objects, $n \propto
S^{-\gamma}$, has an index $\gamma \approx 2.5$. This emphasises the importance
of constraints on the faint-source population, expected to be
dominated by star-forming galaxies, to the overall sky temperature.
To understand how large an effect these sources have we need to consider the
completeness of the surveys. 

\subsection{Completeness}

\begin{figure}
  \includegraphics[width=\linewidth]{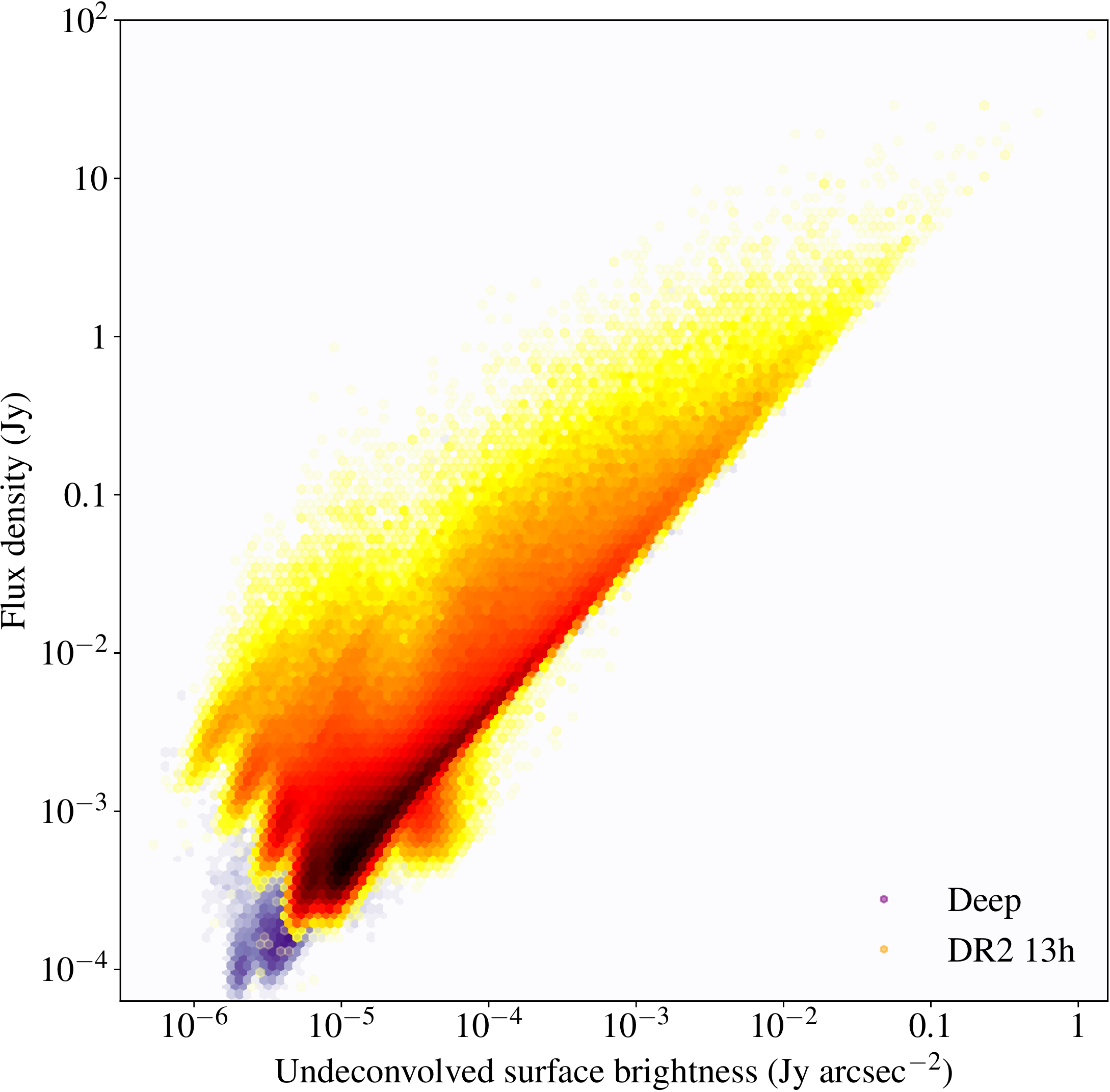}
  \caption{Total flux density against a proxy of mean surface
    brightness for the wide and deep fields. Red/yellow colours show a
    density plot for the 13-h field; purples, below, show the same
    plot for the combination of the three deep fields described in the
    text. The boundary at the lower right-hand edge corresponds to
    unresolved sources, and at the left-hand edge to the surface
    brightness limit.}
  \label{fig:sb}
  \end{figure}

A given source will be catalogued by {\sc PyBDSF} if its {\it peak} flux
density exceeds a certain limiting level relative to the local rms.
This means that LOFAR surveys are not just flux-density limited (with
a flux density limit that depends on position on the sky in various ways) but
also surface-brightness limited in a non-trivial way \citep[as discussed in the context of
  DR1 by][]{Hardcastle+19}. This is
illustrated in Fig.\ \ref{fig:sb}, which shows total flux density against a
proxy of surface brightness (total flux density divided by the product of
major and minor axes) for the 1.9 million
DR2 sources and for the 36,601 sources in the three deep fields.
A survey with a given depth has a minimum flux density, corresponding
to a horizontal line on this plot, but also a complex series of limits
in surface brightness as a result of the multi-scale wavelet strategy
used by {\sc pybdsf}. Sources to the left of the locus of points shown
on this plot are not detectable to the survey. The deeper fields shown
in Fig.\ \ref{fig:sb} clearly have both a lower flux density limit and
a lower surface brightness limit, as expected, but these just
essentially translate the DR2 limits down and to the left without
otherwise modifying them.

The effect of this so-called resolution bias (e.g.
\citealt{Prandoni+01,Williams+16}) is that we need to consider both a
source's total flux density and its angular size in estimating the
completeness of a particular observation; however, some approximation
is needed to the true intrinsic size distribution of sources in order
to do this. For the purposes of this paper, we are particularly
concerned with the possibility that faint diffuse flux may be missed
by the source extraction process, so we require a completeness
correction process that is as sensitive as possible to extended
emission and that estimates the effect on the total sky
  brightness of such incompleteness. These requirements argue against
the use of an assumed angular size distribution derived from
high-frequency data without the short baselines available to LOFAR. In
Fig.\ \ref{fig:sb} we see that DR2 sources with flux densities greater
than a few tens of mJy are unaffected by surface brightness limits
(that is, they essentially all lie well to the right of the series of
surface brightness limits in Fig.\ \ref{fig:sb}). This population is,
of course, expected to be dominated by AGN (or components of AGN) and
so may not be -- indeed, probably is not -- representative of the
population as a whole, but it gives us a completeness estimate which
is directly generated from data comparable to those being considered.
AGN components just at the surface brightness threshold for the survey
are a commonplace sight in the DR2 images and so it is not unrealistic
to imagine that they may have fainter, undetected counterparts. We
emphasise that if, as seems likely, the majority of faint sources are
in fact unresolved star-forming galaxies, this correction will tend to
overestimate the true source counts by a small factor, but we shall
see in later sections that in fact the magnitude of the correction is
not so large as to make a significant difference to the sky
temperature measurement.

Our approach to estimating completeness corrections for the three deep
fields is therefore as follows:
\begin{enumerate}
  \item We start from the existing cropped images and {\sc pybdsf}
    catalogue for each deep field.
  \item For a given total flux density, we insert into the original
    deep-field image Gaussian simulated sources with that total flux density but major and minor axes
    drawn at random from those of the 13-h DR2 population with
    total flux density $>100$ mJy. Since our aim is to assess the
    effect of additional sources on the total recovered flux density,
    these simulated sources are placed in the image completely at
    random with no restrictions; that is, there is no attempt
    to avoid placing a simulated source on top of an existing bright source.
  \item We then re-extract a catalogue from the modified image. In
    general we expect this catalogue to have a higher total flux
    density and a larger number of sources than the original
    catalogue, since it retains all the original sources.
  \item Steps 2 and 3 are repeated many times for each input flux density.
  \item For each input source flux density, the mean ratio, over all
    injected sources and all iterations, of the recovered excess flux
    density to the true input excess flux density gives us the `flux
    density completeness' correction -- the fraction of flux density
    from sources of that flux density that we would expect to be able
    to recover for that field. This, rather than the fraction of sources recovered
    with a given flux density, is the quantity we require for
    estimating corrections to the total integrated flux density from
    all sources.
  \item Steps 2-5 are repeated for a sequence of input flux densities
    in the range 100 $\mu$Jy -- 5 mJy to derive a table of per-field,
    per-flux completeness corrections.
\end{enumerate}

\begin{figure}
  \includegraphics[width=1.0\linewidth]{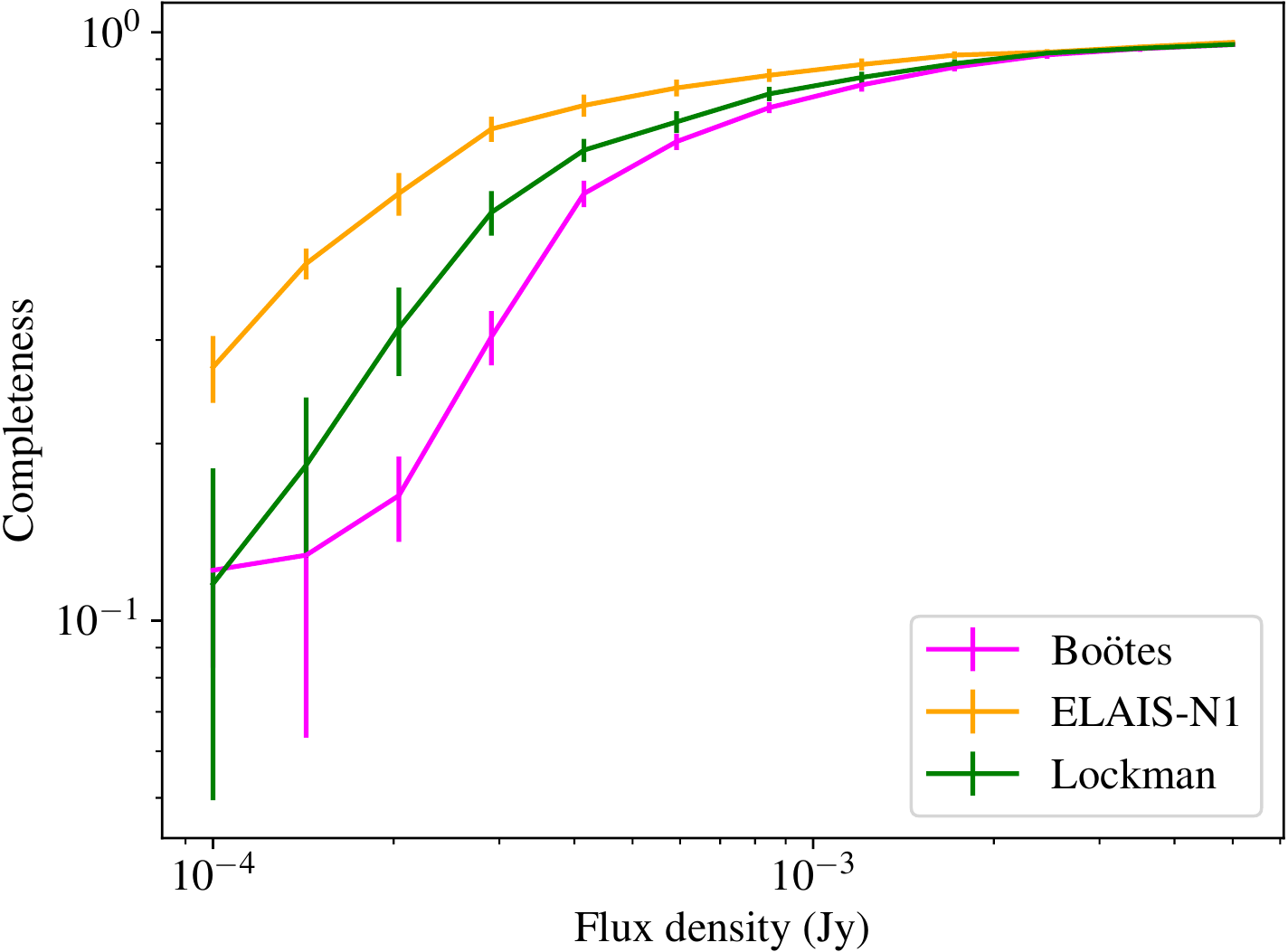}
  \caption{`Flux density completeness' corrections for the three deep
    fields.}
  \label{fig:completeness}
\end{figure}

\begin{figure*}
  \includegraphics[width=0.48\linewidth]{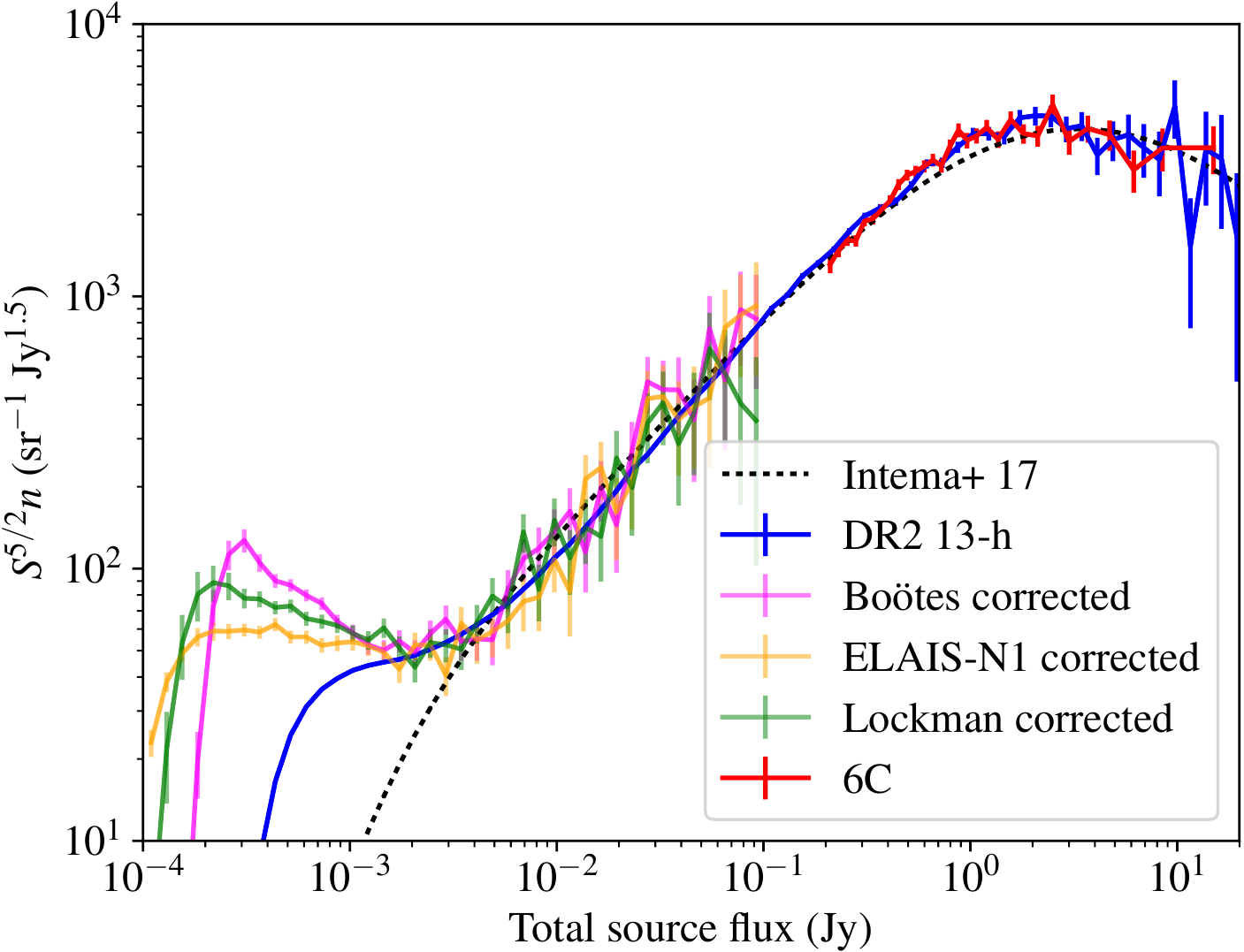}
  \includegraphics[width=0.48\linewidth]{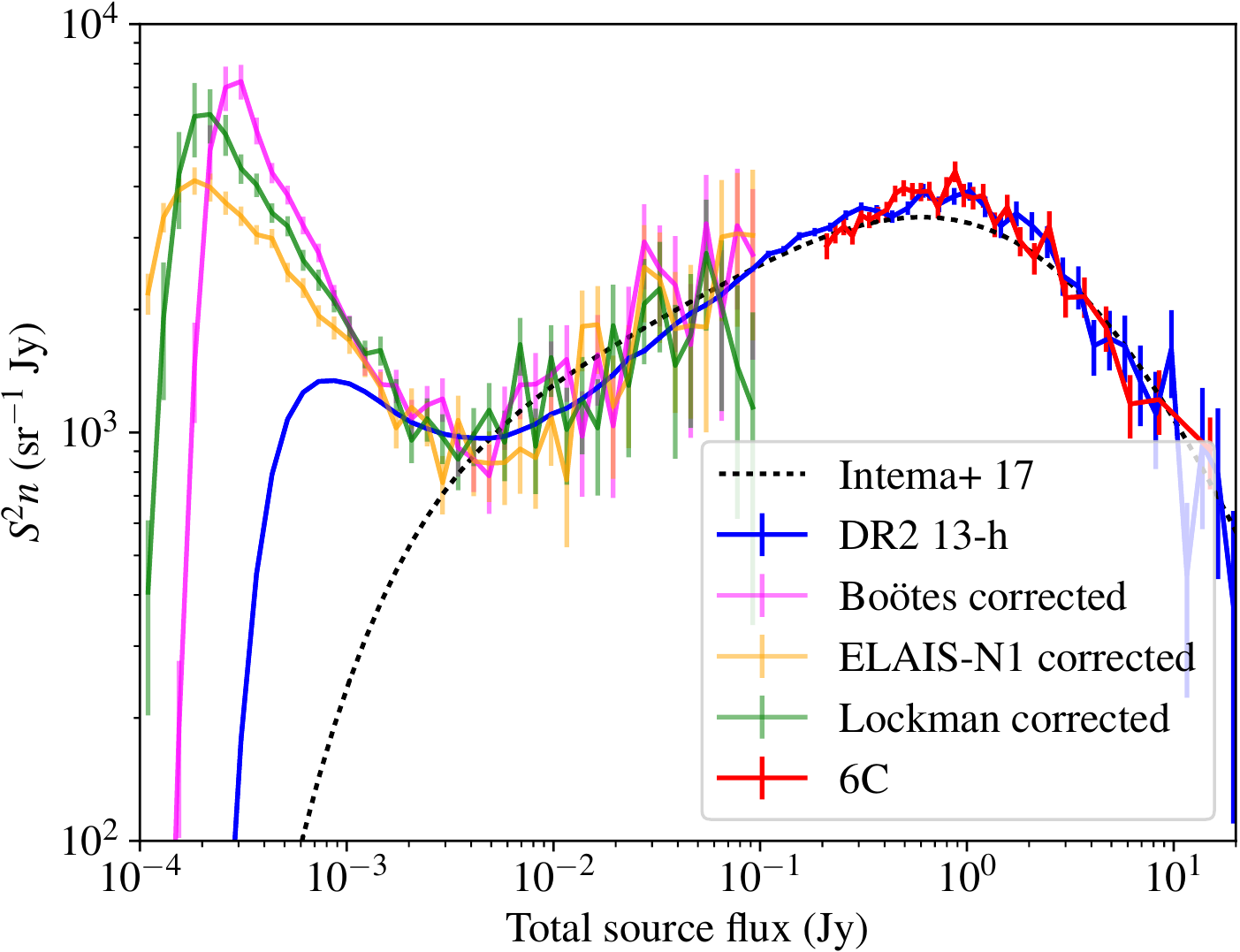}
\caption{Source counts for the deep fields with the `flux-density
  completeness' correction described in the text applied to the data.
  Lines, symbols and other source count plots as in
  Fig.\ \ref{fig:sourcecounts}. Error bars for the deep fields include
  the statistical uncertainties on completeness correction factors.}
\label{fig:c_sourcecounts}
\end{figure*}

Fig.\ \ref{fig:completeness} shows the resulting completeness
corrections, while Fig.\ \ref{fig:c_sourcecounts} shows the source
counts for the three deep fields with these corrections applied -- to
do this we simply linearly interpolate the completeness curves of
Fig.\ \ref{fig:completeness} and apply their reciprocal to the number
count contributed by each source to the binned source counts.
Unsurprisingly, the corrections make a non-negligible difference at
the low flux-density end, implying a power law in numbers that may be
slightly steeper than the Euclidean index of 2.5. The three deep
fields show slightly different corrected curves: the most reliable,
being the deepest, is presumably the ELAIS-N1 result, and there are
large uncertainties on the completeness corrections for the other two
fields below 200 $\mu$Jy. If taken at face
value, the ELAIS-N1 curve implies a genuine turn down in the number
counts of sources below 200 $\mu$Jy at 144 MHz. The other two deep
fields do not constrain this because their completeness corrections
are not reliable below this flux density level. We note that the
implied downturn at the lowest flux densities is already in tension
with some of the models proposed by \cite{Ewall-Wice+20}, which would
imply flat or rising $S^2 n$ below 1 mJy when we combine their sources
with the star-forming galaxy population. However, we are at the limits
of what can be done with the small numbers of sources in our EN1
sample and deeper LOFAR observations are needed to constrain the
population at the lowest flux densities. Mandal \etal\ (in prep.) will
discuss the faint end of the source counts in the deep fields.

\subsection{Total sky temperature}

In principle the sky temperature contribution from a complete sample
is easy to calculate. The spectral radiance $I_\nu$ is given by
\begin{equation}
I_\nu = \frac{\sum S_i}{\Omega}\quad,
\label{eq:inusum}
\end{equation}
where $S_i$ are the flux densities of individual sources in W
Hz$^{-1}$ m$^{-2}$ and $\Omega$ is the total solid angle covered by the
survey in sr. The sky brightness temperature is then given by
\begin{equation}
T_b = \frac{I_\nu c^2}{2k\nu^2}\quad .
\end{equation}
Alternatively, we can integrate over the binned number counts to obtain
\begin{equation}
I_\nu = \int S n(S)\ \rd S\quad ,
\end{equation}
where $n$ is the differential number count, and then convert the
spectral radiance to temperature in the same way.

We are in a position to make the calculation in both ways, since we
can either numerically integrate the binned source counts of
Fig.\ \ref{fig:c_sourcecounts} or add up individual source flux
densities applying the completeness correction to the summed fluxes.
In both cases we switch between the use of the wide-field and
deep-field data at a flux-density value of 4 mJy, conservatively
chosen as the point where the source counts for the wide and deep
surveys diverge; the actual choice of this threshold between 1 and 4
mJy makes very little difference to the results. We verify that the
two possible approaches give the same answer, as expected. For all
deep three fields --- with the DR2 13h data supplying the bright end
of the source counts -- we obtain a total spectral radiance (sky surface
brightness) of approximately 28 kJy sr$^{-1}$ at 144 MHz. Using integration of
the source counts down to the lowest levels for which we have reliable
completeness corrections, we then obtain sky temperature values of $44.5 \pm 3.0$
K for Bo\"otes, $43.3 \pm 2.8$ K for ELAIS-N1 and $45.4 \pm 3.3$ K for
Lockman, where the error estimates take account of the flux scale
uncertainties that we have assigned to the wide and deep fields (the dominant contribution)
but also the statistical uncertainties on the completeness correction factor.
Taking the mean of the three fields, our best estimate for the total
contribution to the sky temperature of discrete sources above 100
$\mu$Jy is $44.4 \pm 1.7$ K. We note that the effect of the flux
density completeness correction is generally a boost to sky brightness
or temperature by between 10 and 20 per cent, but only of order 10 per
cent for our most reliable field, ELAIS-N1. Thus, even if our
completeness corrections are over-generous as discussed above, a
conservative lower limit on the contribution of discrete sources to
the sky temperature would be $\sim 40$ K.

\subsection{Consistency check using images}
\label{sec:imagecheck} 
For the DR2 data we carried out a further consistency check to
establish whether significant extended flux density was missed by the
PyBDSF decomposition itself. To do this we took the mosaic images
covered by our catalogue, both at high (6 arcsec) and low (20 arcsec)
resolution, calculated the number of non-blank pixels and the sum of
the flux density over those pixels, and so calculated a spectral
radiance for each field. The mean of these values is then the mean sky
brightness contributed by sources above the noise level in the
wide-field data. (Because the mosaics overlap, each sky pixel is not
uniformly weighted in this sum, but no bias shoud be introduced by
this process.) We emphasise that it is not safe to sum over the maps
alone to obtain the sky temperature, because any undeconvolved emission
present will not be accounted for correctly, but the sums can be used
to check whether PyBDSF is missing flux. Summing over $2.1\times
10^{11}$ pixels in the full-resolution mosaics, we obtain a mean of
$21.8 \pm 0.2$ kJy sr$^{-1}$ (statistical uncertainties only), which
is in excellent agreement with the value obtained by integration of
the source counts for these fields (21.8 kJy sr$^{-1}$) or by direct
summation of the source fluxes (21.7 kJy sr$^{-1}$). Interestingly,
the sum over the low-resolution mosaics is slightly higher at $22.5
\pm 0.2$ kJy sr$^{-1}$, which could reflect either a tendency for
diffuse structure to be missed in the maps and catalogues at 6 arcsec
resolution or some differential calibration uncertainty between long
and short baselines. The difference is only a few per cent and thus
does not have a significant effect on our conclusions.

\section{Discussion and conclusions}

We have estimated the total contribution of discrete sources between
100 $\mu$Jy and 100 Jy to the sky temperature at 144 MHz to be $44 \pm
2$ K. We derived this from a combination of three deep surveys,
extending down to an rms noise level of 20 $\mu$Jy per beam, after
applying completeness corrections and deriving the bright end of the
source counts from a shallower but much larger survey conducted with
the same instrument and analysed using comparable techniques. Our
estimate should thus include contributions from discrete sources with
144-MHz flux densities from 100 $\mu$Jy to $\sim 100$ Jy.

Our estimate of the sky temperature from discrete sources is substantially higher than the best existing discrete
source-count based estimate at these frequencies given by
\cite{Vernstrom+11}, who used only the 6C counts to obtain $T = 18$ K
at 150 MHz (20 K at 144 MHz).\footnote{Here and throughout where a
  temperature index is not known we convert quoted literature
  temperatures to 144 MHz on the assumption of a temperature index
  $\beta = 2.7$, $T \propto \nu^{-\beta}$, since the spectral index of
  the extragalactic population is known to be $\alpha \approx 0.7$, $S
  \propto \nu^{-\alpha}$: see, for example, \cite{Hardcastle+16}.} Our value exceeds even their highest
extrapolation down to low flux densities, presumably because of the
non-negligible contribution from both faint and diffuse sources that
were not present in their extrapolation.

Our result is in remarkable agreement with the estimate of
\cite{Bridle67} of a sky temperature for the isotropic component of
$30 \pm 7$ K at 178 MHz ($53 \pm 12$ K at 144 MHz) and of course is
consistent with the rough upper limit on temperature from the northern-sky
survey of \cite{Turtle+Baldwin62} (80 K at 178 MHz or 140 K at 144 MHz).

However, it is clear that the LOFAR-detected source population cannot
explain the entirety of the background reported by \cite{Fixsen+11},
which would correspond to 190 K at 144 MHz (we use the best-fitting
power-law model from their work) and which is supported by more recent
independent observations such as those of \cite{Dowell+Taylor18}. Even
with LOFAR's sensitivity to both faint and extended sources, we are
reproducing only a quarter of the ARCADE-2 background level. If the
downturn in the corrected source counts below 200 $\mu$Jy is to be
believed (and it seems qualitatively consistent with what is expected
from source count models and $P(D)$ analysis at higher frequencies:
\citealt{Condon+12,Mauch+20}) then no extrapolation of the currently
detectable source population will make up this discrepancy; once the
number counts fall below $n \propto S^{-2}$ they rapidly cease to make
a significant contribution to the sky temperature. If we ignore any
evidence for a downturn, then power-law extrapolation, $n \propto
S^{-5/2}$, of the observed number counts from their observed peak
level at around 200 $\mu$Jy down to levels of a few $\mu$Jy would
explain all of the ARCADE-2 background, but this is very definitely
not what is expected from source models, and \cite{Condon+12} argue
that such a numerous bright source population is ruled out by $P(D)$
analysis; in any case the long baselines of LOFAR would be needed to
avoid the confusion limit in studying such a source population down to
the lowest flux levels.

The use of LOFAR, with its excellent sensitivity to extended emission,
also rules out the possibility that a significant fraction of the
excess can be due to diffuse radio emission, as discussed by, for example,
\cite{Vernstrom+15}, unless it is well below the sensitivity limits of
even the deep surveys. The fact that we can directly account for a
higher fraction of the ARCADE-2 background than has hitherto been
possible may be a result of LOFAR's excellent sensitivity to resolved
sources, but if our surface brightness completeness analysis is valid
then we have taken this as far as it can go. Only if a population of
faint sources existed whose sizes were much larger than can be
extrapolated from the $>100$ mJy population (i.e., probably many arcmin
in size) would we be significantly
in error. Such sources would likely be modelled out in our calibration
and imaging process and a dedicated low-resolution calibration and
imaging of the
LOFAR data might be necessary to search for them. The conclusions of
Section \ref{sec:imagecheck} suggest that the resolution would need to
be of the order of arcminutes to have any chance of revealing flux
missed by the current analysis, but LOFAR surveys
data tapered to arcmin resolution have significantly reduced
sensitivity and so cannot be used to investigate this further.

The level of the ARCADE-2 excess is very sensitive to a model of the
Galactic foreground emission; the analysis of \cite{Fixsen+11} and in
particular the Galactic modelling of \cite{Kogut+11} were carried out
before the most recent {\it Planck} results were available, which show
a strong contribution from anomalous microwave emission (AME) in the
ARCADE-2 band \citep{Planck16}; in addition, there is some evidence
from more recent work, for example with C-BASS\footnote{The C-Band All
  Sky Survey.}\citep{Dickinson+19} that the
spectral index of Galactic emission is steeper than the modelling of
\cite{Kogut+11} would allow for. These factors might make the
remaining extragalactic excess -- which must exist in the data --
more compatible with the levels found here and elsewhere from discrete
sources.

A remaining possibility is that the ARCADE-2 excess is the result of a
previously unaccounted for synchrotron background that has most of its
power on such large angular scales ($\sim 1$ degree) such that it is
largely undetectable by the LOFAR observations we have used. It is
hard to see how such a background could be related to a known
extragalactic source population. Krause \& Hardcastle (in prep.) will
discuss a possible model in terms of emission from sources very local
to the solar system (the Local Bubble).

Our findings are also relevant to the search for signals from cosmic
dawn, in particular to the modelling of the foreground contributions
to the sky-averaged spectrum. EDGES \citep{Bowman+18} reported a
deficit in the extragalactic brightness temperature due to HI
absorption of about 0.5 K at a frequency around 78 MHz. In their
analysis they model ionospheric and galactic foregrounds (see also the
discussion by \citealt{Hills+18} and \citealt{Bowman+18b}), but do not
address diffuse extragalactic foreground components explicitly.
However, their five-parameter foreground model accounts implicitly (at
least partially) for extragalactic synchrotron foregrounds as well.
The steepest foreground component in their model goes as $T_\nu
\propto \nu^{-2.5}$, whereas, as noted above, we would expect the
extragalactic background to have a steeper spectrum. 
Extrapolation of our measurement at 144 MHz by means of a power-law
spectrum to the frequencies probed by EDGES would imply an extragalactic
brightness temperature at 78 MHz of $T \approx 235 \pm 18$ K,
allowing for a spectral index uncertainty of $\pm 0.1$. In contrast to
the Galactic and ionospheric foregrounds, the extragalactic component
should be stationary and isotropic, modulated by a cosmic dipole; its
magnitude would also be comparable to the claimed absorption
signal. Deep fields observed with the LOFAR LBA will allow us in the future
to directly probe the relevant frequency window.

Although the LOFAR deep fields used in this work represent the deepest
images ever made at 100-200 MHz frequencies, it is possible to go
still deeper -- additional data on these fields and on others (for example,
the North Celestial Pole field) exist and, on a timescale of a few
years, it will be possible to improve our estimates of the
contribution to the sky temperature of the $<100$ $\mu$Jy population.

\section*{Acknowledgments}

MJH would like to acknowledge conversations with Tessa Vernstrom which
inspired him to make use of LOFAR data to address this science
question. We are grateful to G\"ulay G\"urkan and Alexandar Shulevksi
for comments on the first draft of the paper.

MJH acknowledges support from the UK Science and Technology
Facilities Council (ST/R000905/1).  PNB and JS are grateful for support from the UK STFC via grant ST/R000972/1.  AD acknowledges support by the
BMBF Verbundforschung under the grant 05A17STA. MJJ acknowledges
support from the UK Science and Technology Facilities Council
[ST/N000919/1] and the Oxford Hintze Centre for Astrophysical Surveys
which is funded through generous support from the Hintze Family
Charitable Foundation.   IP acknowledges support from INAF under the SKA/CTA PRIN “FORECaST” and the PRIN MAIN STREAM “SAuROS” projects.

LOFAR, the Low Frequency Array designed and constructed by ASTRON, has
facilities in several countries, which are owned by various parties
(each with their own funding sources), and are collectively operated
by the International LOFAR Telescope (ILT) foundation under a joint
scientific policy. The ILT resources have benefited from the
following recent major funding sources: CNRS-INSU, Observatoire de
Paris and Universit\'e d'Orl\'eans, France; BMBF, MIWF-NRW, MPG, Germany;
Science Foundation Ireland (SFI), Department of Business, Enterprise
and Innovation (DBEI), Ireland; NWO, The Netherlands; the Science and
Technology Facilities Council, UK; Ministry of Science and Higher
Education, Poland.

Part of this work was carried out on the Dutch national
e-infrastructure with the support of the SURF Cooperative through
grant e-infra 160022 \& 160152. The LOFAR software and dedicated
reduction packages on \url{https://github.com/apmechev/GRID_LRT} were
deployed on the e-infrastructure by the LOFAR e-infragroup, consisting
of J.\ B.\ R.\ Oonk (ASTRON \& Leiden Observatory), A.\ P.\ Mechev (Leiden
Observatory) and T. Shimwell (ASTRON) with support from N.\ Danezi
(SURFsara) and C.\ Schrijvers (SURFsara). This research has made use of the University
of Hertfordshire high-performance computing facility
(\url{https://uhhpc.herts.ac.uk/}) and the LOFAR-UK compute facility,
located at the University of Hertfordshire and supported by STFC
[ST/P000096/1]. The J\"ulich LOFAR Long Term Archive and the German
LOFAR network are both coordinated and operated by the J\"ulich
Supercomputing Centre (JSC), and computing resources on the
supercomputer JUWELS at JSC were provided by the Gauss Centre for
supercomputing e.V. (grant CHTB00) through the John von Neumann
Institute for Computing (NIC). This research made use of {\sc Astropy}, a
community-developed core Python package for astronomy
\citep{AstropyCollaboration13} hosted at
\url{http://www.astropy.org/}, of {\sc Matplotlib} \citep{Hunter07},
and of {\sc topcat} \citep{Taylor05}.
\clearpage
\bibliographystyle{aa}
\renewcommand{\refname}{REFERENCES}
\bibliography{../bib/mjh,../bib/cards}

\end{document}